\documentclass[10p, twocolumn]{revtex4-2}

\usepackage{mathrsfs}
\usepackage{amsfonts}
\usepackage{amssymb}
\usepackage{amsmath}
\usepackage{graphicx}
\usepackage{color}
\usepackage{stackrel}
\usepackage[utf8]{inputenc}
\usepackage{amsmath, amssymb, amsfonts}
\usepackage{geometry}
\usepackage{booktabs}
\usepackage{array}
\usepackage{hyperref}

\newcommand{\be}{\begin{equation}}
\newcommand{\ee}{\end{equation}}
\newcommand{\bqa}{\begin{eqnarray}}
\newcommand{\eqa}{\end{eqnarray}}

\begin{document}

\title{Emergent Entropic Forces Drive Inverse and Reentrant Phase Transitions 
in Condensates of Intrinsically Disordered Proteins}

\author{D. A. Herrera-Mart\'i}
\affiliation{Centre for Quantum Technologies, National University of Singapore, 3 Science Drive 2, Singapore 117543, Singapore}
\affiliation{CEA List, Universit\'e Grenoble Alpes, France}

\author{Marc Jamin}
\affiliation{Institut de Biologie Structurale (IBS), Univ. Grenoble Alpes, CEA, CNRS,  38000 Grenoble, France}

\date{\today}

\begin{abstract}
 
We present a phenomenological, mean-field model for condensates of intrinsically disordered proteins that describes inverse phase transitions and qualitatively predicts reentrant phase behavior as a function of salt concentration. The model is based on a matrix-based formulation of the Flory-Huggins free energy, in which the enthalpic interaction parameter and the protein valence, defined here as the number of effective ``sticky patches”, are treated as free parameters. Its central insight is that integrating out dynamics of fast-moving molecular species generates an effective entropic force that increases the free-energy barrier to mixing, even within a mean-field description. We account for this by linking Gaussian integration of co-solutes to the Schur complement of the matrix-based Flory-Huggins free energy. This provides a general thermodynamic framework illustrating how enthalpic interactions that increase the number of available microstates necessarily induce effective entropic forces. We demonstrate the expressiveness and predictive power of the model by qualitatively reproducing recent experimental observations on the rabies virus phosphoprotein.
\end{abstract}

\maketitle

\section{Introduction}

Liquid-liquid phase separation (LLPS) is a fundamental physical process whereby biomolecules, especially proteins and RNAs, can demix from the surrounding cellular environment to form dynamic, membrane-less compartments \cite{holehouse24, alberti19, uversky18,dignon20,banani20}. These phase transitions can be triggered by changes in intensive solution parameters such as temperature, pH, ionic strength, or molecular concentration. Temperature-dependent phase separation is particularly well characterized in terms of the balance between entropic and enthalpic contributions to the free energy of mixing. In upper critical solution temperature (UCST)-type systems, cooling promotes phase separation because the balance between enthalpic and entropic contributions to the free energy shifts in favour of demixing. Above the UCST, the different components of the mixture can minimize the free energy beyond entropic effects by  mixing with each other. Below the UCST, the components can instead minimize the free energy by reducing their mutual contacts and separating into distinct phases, as the enthalpic interactions favouring demixing overcome the entropic gain of mixing. The opposite behavior, inverse phase transition, also referred to as lower critical solution temperature (LCST)-type LLPS, describes a phenomenon in which phase separation is promoted upon increasing temperature. In contrast to UCST systems, LCST-type phase separation in biomolecular systems emerges from the temperature-dependent balance between protein hydration, solvent entropy, and polymer conformational changes  \cite{falahati19,dignon20,bari21}. Different intrinsically disordered polypeptides, including elastin-like proteins, exhibit inverse phase transitions \cite{meyer04, macewan14}, making LCST behavior highly relevant for biomolecular condensation and intracellular organization.  Reentrant phase transitions (RPTs) are characterized by the reappearance of a thermodynamic phase upon the monotonic variation of a single control parameter. For example, a homogeneous solution may undergo LLPS upon heating or cooling and subsequently remix into a single homogeneous phase as the same parameter continues to monotonically vary. In temperature-driven systems, reentrant behavior is characterized by the successive occurrence of a lower critical solution temperature (LCST) transition and an upper critical solution temperature (UCST) transition, or vice versa, resulting in a closed-loop coexistence region in the temperature–composition phase diagram.

Intrinsically disordered proteins (IDPs) are remarkably prevalent in the eukaryotic proteome, where approximately 20–30\% of proteins contain long intrinsically disordered regions (IDRs) \cite{ward04,peng15,wright15, vdlee14}. Their structural plasticity enables multivalent, dynamic and transient interactions, making them central regulators of signaling, transcription, and biomolecular condensation \cite{holehouse24}. Intrinsic disorder is also extensively exploited by viruses, whose proteins often engage multiple host partners, evade immune responses, and orchestrate the assembly of viral replication complexes through flexible interaction interfaces \cite{xue10,lihaohua22}. Beyond their essential physiological roles, dysregulation or aberrant behavior of IDPs has been implicated in numerous human diseases, including neurodegenerative disorders such as Alzheimer’s and Parkinson’s diseases \cite{rai21}. 

IDPs can undergo LLPS to form dynamic biomolecular condensates in cells \cite{holehouse24, uversky18}. These condensates constitute transient, reversible non-covalent molecular assemblies in which individual protein molecules associate through a large number of weak intermolecular interactions, including electrostatic and hydrophobic interactions, hydrogen bonding, $\pi$–$\pi$ and cation–$\pi$ interactions  \cite{holehouse24,alberti19}

The Flory-Huggins (FH) model for polymer solutions \cite{flory42, rubinstein03,qian22, souza24, choi20} provides a useful mean-field framework for describing such biological systems but fails to capture many defining features of biomolecular condensates. In particular, it treats polymers as chemically homogeneous chains interacting through averaged pairwise interactions, thereby neglecting sequence-specific effects, heterogeneous charge patterning, transient secondary structure, multivalent binding and the nonequilibrium nature of the cellular environment. Several theoretical extensions, including Debye-H\"uckel, Poisson-Boltzmann electrostatics and Random Phase Approximation (RPA) approaches \cite{li22,harries13,lin17} have substantially improved the description of charged IDPs. However, these increasingly detailed models rapidly become computationally demanding, making them difficult to apply to the large, multicomponent systems characteristic of cellular condensates. Moreover, because biomolecular condensates are inherently coarse-grained objects in which microscopic interactions are effectively averaged over binding sites, stickers, and molecular conformations, increasing microscopic detail does not necessarily translate into improved predictive power at the mesoscopic scale. These considerations motivate the development of a phenomenological mean-field framework that retains the essential microscopic physics while remaining sufficiently simple to describe the thermodynamics of complex biomolecular condensates.

Here, we introduce a phenomenological extension of FH theory that naturally captures LCST-type behavior through a small set of physically motivated effective parameters. The framework provides a simple mean-field description of entropy-driven phase separation while remaining sufficiently flexible to account for the influence of additional molecular species, such as salts or co-solutes. We assess the predictive power of the model by comparing its phase behavior with recent experimental observations of the rabies virus phosphoprotein \cite{bouchama25},which exhibits an LCST transition and monovalent salt-dependent reentrant phase behavior. Despite its simplicity, the model qualitatively reproduces the experimentally observed phase diagram and provides a statistical-mechanical interpretation of the underlying thermodynamic mechanisms.

\section{Theoretical Framework}

The classical thermodynamic descriptions of macromolecular solutions rely fundamentally on FH theory, which can account for the substantial asymmetry in conformational entropy between high-molecular-weight polymers and small solvent molecules, based on a per-site, mean-field approach. The dimensionless free energy of mixing per lattice site, $f$, for a binary polymer-solvent system is:

\be
\label{eq:flory_huggins}
\frac{f}{k_B T} =  \underbrace{\frac{\phi}{N} \ln \phi + (1-\phi) \ln (1-\phi)}_{\textrm{Entropic term}} + \underbrace{\chi \phi (1-\phi)}_{\textrm{Enthalpic term}}
\ee
where $\phi \in [0,1]$ denotes the volume fraction of the macromolecule, $N$ represents the degree of polymerization (the number of lattice sites occupied by a single polymer chain, with $N=1$ for a monomeric solvent), and $\chi$ is the dimensionless Flory-Huggins interaction parameter. The parameter $\chi = 2 u_{AB} - (u_{AA} + u_{BB})$ captures the energetic differential between homotypic and heterotypic contacts and should be undestood as a free energy term. One should think about $u_{ij}$ as the energetic barrier to be overcome for to species $i$ and $j$ to share the same site. If it is positive, it is an enthalpic penalty to be surmounted by thermodynamical processes, whereas if its negative, it contributes to the entropic mixing and enforces the mixture. 

The mixing and demixing landscapes are governed entirely by the convexity of $f(\phi)$ as tuned by $\chi$. When $\chi$ exceeds a critical threshold $\chi_c$, the free energy curve develops two distinct local minima. This gives rise to phase separation and divides the phase diagram into three thermodynamic regions bounded by the binodal and spinodal curves \cite{rubinstein03}.  The binodal defines the compositions of the two coexisting phases in macroscopic thermodynamic equilibrium. It is obtained from the conditions of equal chemical potential and equal osmotic pressure (or equivalently, equal tangent construction on the free-energy curve) between the two phases. In contrast, the spinodal marks the limit of local thermodynamic stability. It corresponds to the locus of composition at which the homogenous phase becomes unstable with respect to infinitesimal concentration fluctuations. Mathematically, it is defined by the inflection point of the free energy density, where its second derivative with respect to the composition vanishes:

\be
\label{eq:spinodal_condition}
\frac{\partial^2 f}{\partial \phi^2} = 0
\ee

\begin{figure}
\includegraphics[width=\linewidth]{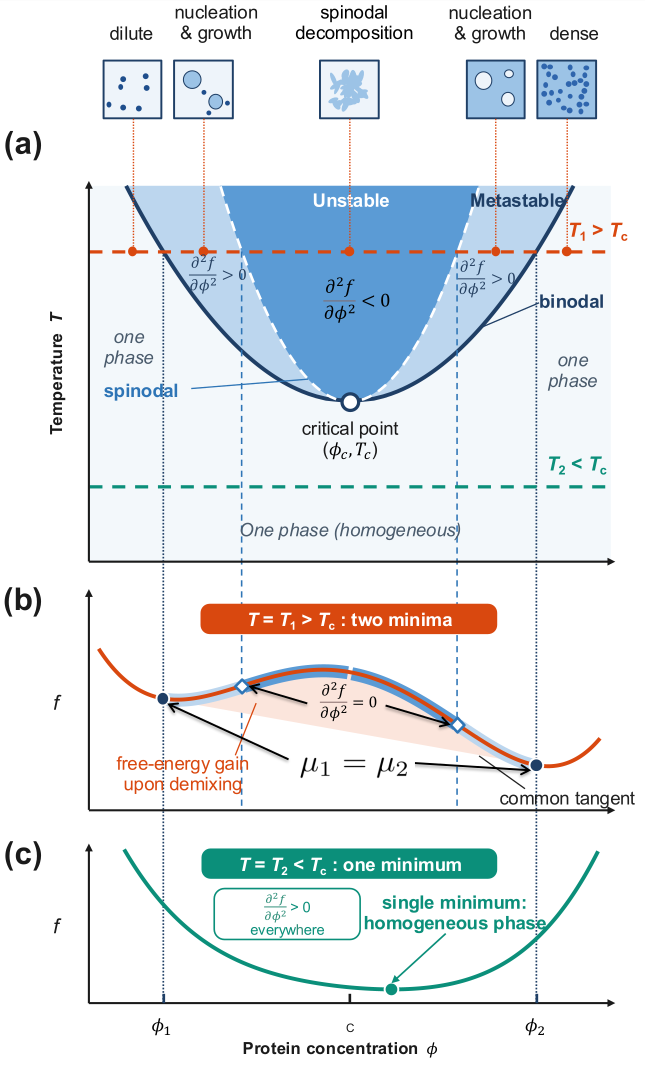}
  \caption{The theory of Flory-Huggins predicts a free energy per lattice which is a combination between an entropic term (i.e. a binary entropy potential facing down) plus an enthalpic term (which can be concave or convex). {\bf (a)} Temperature-concentration phase diagram illustrating an LCST transition. The solid blue (\emph{binodal}) defines the coexistence boundary, while the dashed red \emph{spinodal} marks the limit of local thermodynamic stability ($\partial^2 f / \partial \phi^2 = 0 $), meeting at the critical point ( $T_c, \phi_c$). Isotherms $T_1 > T_c$ and $T_2 < T_c$ highlight the two-phase demixed and single-phase homogeneous regimes, respectively. {\bf (b)} For some values of temperature, on-convex double-well landscape yields two coexisting compositions, corresponding to different concentrations of polymers.chemical potential, computed as the tangent($\mu_1 = \partial f / \partial \phi|_{\phi_1} = \partial f / \partial \phi |_{\phi_2} = \mu_2$), must be the same at both concentrations. Similarly, the osmotic pressure associated to both minima ($\Pi = -f + \phi\mu$) must be the same for the solution to be locally stable. The locus defined this way for different temperatures is known as the \emph{binodal}. At the free energy minimum, each real-space point in the solution ``chooses" a given concentration. In the unstable phase, there is no well-defined configuration and the concentration at each point fluctuates. {\bf (c)} Free-energy density curve below the critical temperature ($T_2 < T_c$). Entropic and enthalpic forces collaborate rather than compete and the curve is strictly convex with a single global minimum, stabilising the homogeneous mixed state.} 
\label{lcst} 
\end{figure}

The region bounded between the binodal and spinodal curves ($\partial^2 f / \partial \phi^2 \gtrapprox 0$) is metastable. Conversely, the region enclosed within the spinodal curve ($\partial^2 f / \partial \phi^2 < 0$) is inherently unstable (Fig.~\ref{lcst}(a)).  Here, infinitesimal concentration fluctuations spontaneously grow without encountering a free-energy barrier, resulting in spontaneous phase separation through spinodal decomposition.

\subsection{Limitations of FH: The Lower Critical Solution Temperature}

A critical limitation of the original FH theory is that it intrinsically predicts only UCST-type behavior and cannot account for LCST phase separation without introducing additional temperature-dependent contributions or empirical modifications to the interaction parameter. In the original FH theory,  is derived from mean-field contact energies and is approximately proportional to $1/T$, i.e.~$d \chi / d T < 0$. Thus, $\chi$ decreases   as temperature increases, while the entropy of mixing is purely combinatorial and therefore always favours mixing. As temperature increases, the entropic contribution increasingly dominates the free energy of mixing, stabilizing the homogeneous phase. Consequently, phase separation is predicted to occur only upon cooling, giving rise to UCST-type behaviour. 

In LCST-type behavior, the systems exist as a single homogeneous phase at low temperatures but undergoes phase separation upon heating, corresponding to an increase in the interaction parameter with temperature: $d \chi / d T > 0$. LCST behavior is therefore fundamentally driven by non-combinatorial, solvation entropy and directional non-covalent interactions (such as hydrogen bonding networks). At low temperatures, the free energy of hydration reflects contributions from both polar and hydrophobic surface areas. The hydration of non-polar groups depends on excluded-volume effects, as described by scaled-particle theory, together with the energetic cost of perturbing the surrounding hydrogen-bond network \cite{wolfenden78}. In contrast, polar groups are stabilized by direct hydrogen bonding with water \cite{graziano05}. In addition, van der Waals interactions between water and non-polar surfaces can enhance the local hydrogen-bond network, further influencing the temperature dependence of protein solvation \cite{vegt17}. As temperature increases, the thermodynamics of hydration evolve. The favourable hydration of non-polar surfaces diminishes as excluded-volume contributions become increasingly costly and the stabilization provided by the surrounding hydrogen-bond network is reduced. Consequently, the free-energy penalty for exposing hydrophobic surface area to water increases, making intermolecular association progressively more favourable. When this temperature-dependent change in solvation free energy outweighs the loss of conformational and translational entropy accompanying condensation, phase separation becomes thermodynamically favourable.

Within the FH framework, this behavior cannot be captured by treating the interaction parameter  as arising solely from energetic interactions, for which $\chi \propto 1/T$. Instead, reproducing LCST-type behavior requires an effective interaction parameter that increases with temperature: $d \chi / d T > 0$, reflecting the growing free-energy cost of mixing. Several phenomenological extensions have been proposed \cite{mumby93,chee91,baulin02}, such as simply introducing a negative B coefficient in the temperature dependence of :

\be
\chi = A \pm \frac{B}{T}
\ee
where A and B are constants (see Fig.~\ref{mixingbarrier}(a)). Although these approaches can reproduce LCST phase diagrams, they remain largely empirical and are not derived from a microscopic description of protein solutions.

\subsection{Matrix-Based Flory-Huggins With LCST and Reentrant Phase Behaviour} 

Another important limitation of classical FH theory arises from its mean-field nature. By replacing microscopic interactions with an average interaction field, the theory neglects spatial correlations and local fluctuations in molecular organization. In addition, intermolecular contacts are treated as isotropic and effectively unlimited, implicitly assuming that each molecule can form an infinite number of favourable interactions. Real proteins and polymers, however, possess only a finite number of interaction sites (finite valency) whose binding capacity becomes saturated. This finite valency constrains network connectivity and introduces an additional configurational entropy associated with the multitude of possible bonding topologies, particularly near the percolation threshold. Such correlation and finite-valency effects can profoundly influence phase behaviour but are not captured by the original FH framework \cite{choi20}. However, particles and polymers have finite valency and their bonds can saturate, which leads to maximisation of entropy due to graph percolation \cite{stockmayer43}. 

The core of our contribution is an extension of FH theory that naturally captures the emergence of LCST behavior and reentrant phase separation within a phenomenological yet predictive framework. It incorporates concepts of lattice modes, for instance by turning N into a free parameter that accounts for effective size /finite valency rather than for degree of polymerisation (as is the case in FH)  \cite{borcherds21, choi19} Inspired by modern lattice models \cite{choi19}, our formulation generalizes the classical description by treating the effective lattice occupancy of the polymer as a free parameter. Rather than reflecting only the polymer size, this parameter also accounts for the finite valency of intermolecular interactions, thereby incorporating the effects of saturable bonding into the thermodynamic description  \cite{borcherds21, choi19}. Building on concepts from generalized lattice models, we introduce an effective polymer weight as a free parameter. Unlike the classical FH theory, where this quantity is determined solely by polymer size, our formulation allows it to capture additional entropic contributions arising from finite-valency, saturable intermolecular interactions  \cite{borcherds21, choi19, dignon18}. Given that our model is averaged by site (that is, mean-field), which is necessary to perform integrations that will be explained below, we can only incorporate insights from lattice models in the parameters $N$ and $\chi$ in an implicit manner.

Whereas FH was initially conceived to study the interplay between entropy and enthalpic forces of polymers of fixed length in a solution, organelles and coacervates (either in the cytoplasm or in vitro solutions) consist frequently of fuzzy complexes, which can be dynamic \cite{holehouse24}. It has indeed been suggested that cell homeostasis depends heavily on multicomponent LLPS  \cite{deviri21}. This means that the components, be it IDPs, RNA, combinations thereof or other scaffolds are continuously associating and dissociating to form multimers that come in and out of existence. Thus, the main building blocks are colloidal particles of a given weight that associate without a fixed stochiometry, and therefore cannot be described by the canonical polymer length parameter $N$. This is because coordination valency is determined not by polymer length (RNA proteins are thousands of times heavier than a single water molecule) but by the ratio of ``sticky patches" in the molecule, which dramatically reduces the $N$ in the solution (one can safely assume that the rest of the macromolecule displaces water and reduces the effective solution volume). 

The language of matricial FH for several species can be modified to describe the discussed scenario. A minimal model for this picture would be:

\be
\frac{f}{k_B T} = \sum_i \frac{\phi_i}{N_i} \ln \phi_i + \sum_{ij} \phi_i \chi_{ij} \phi_j
\ee
where each component $\phi_i$ corresponds to a different species in the solution, including solvent, ligands, scaffolds, salts, crowding agents ..., each with effective $N_i$, which, as exlpained above, is a free parameter that measures the number and does not necessarily represent the polymer length. The concentrations are normalised, $\sum_i \phi_i = 1$. We assume that the enthalpic interaction

\be
\chi_{ij} = A_{ij} + \frac{B_{ij}}{T}
\label{eqchi}
\ee
fluctuates randomly around some nominal value (as a result from stochastic drift, screening etc). The multicomponent mixture described by this matricial FH model becomes unstable when the Hessian ($H_{ij} = \partial^2 f/\partial x_i \partial x_j$) of the per-site free energy features at least one negative eigenvalue. The spindal condition is therefore (see Appendix):

\be
\det  H = 0
\ee

Physically, a negative eigenvalue entails the reduction of the associated mode in the distribution. Mathematically this is the hallmark of a saddle point that allows for very rich phase diagrams.

\subsubsection{Integrating out fast degrees of freedom: Schur's Complement}

Our system can be described in terms of {\bf slow degrees of freedom}, associated with the protein scaffold and its collective organization, and {\bf fast degrees of freedom}, corresponding to small molecules such as salt ions and solvent molecules that equilibrate on much shorter timescales. Because the fast degrees of freedom rapidly adjust to any change in the slow protein variables, they can be integrated out to obtain an effective free-energy description of the protein system alone. In the Gaussian approximation, this elimination leads naturally to the {\bf Schur complement}, which accounts for the renormalisation of the protein–protein interactions induced by the fluctuations of the fast degrees of freedom.

\begin{figure}[t!]
\includegraphics[width=\linewidth]{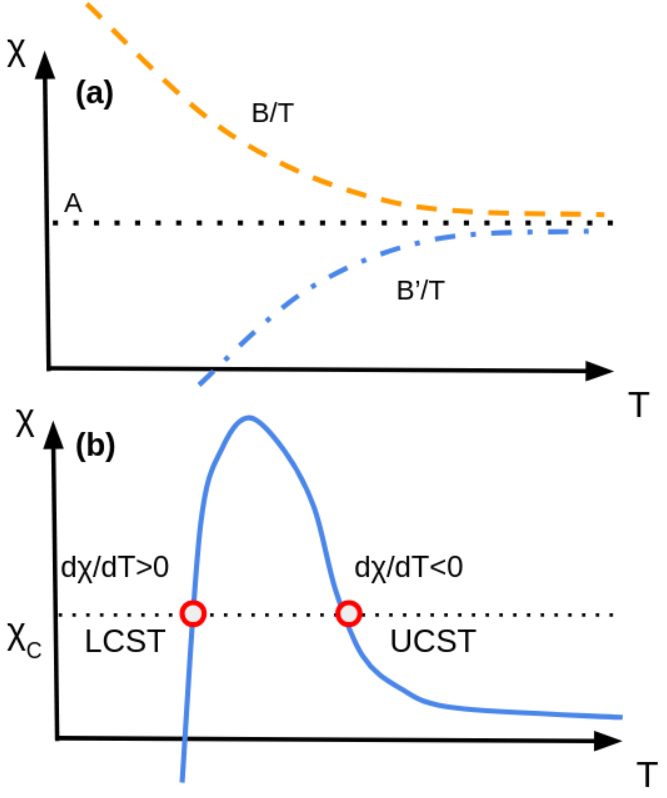}
\caption{Intuition on the mixing barrier and how to obtain a reentrant. {\bf (a)} Usually, the Flory-Huggins parameter is decomposed into an entropic (A) and an enthalpic term (B), $\chi = A + B/T$. Flory-Huggins only admits positive A and B coefficients although it is commonplace to assume the form $\chi' = A - B'/T$, again for $B'$ positive, in order to account for LCTS. These form for the FH parameter converge to $A$ in the high temperature limit. {\bf (b)} Conceptually, our derivation amounts to a mixing barrier that increases at low temperatures, and can lead under the right enthalpic interactions, to either a bubble or hourglass reentrant mechanism. This is due to the mediated interaction by the fast degrees of freedom. } 
\label{mixingbarrier} 
\end{figure}

The matrix-based FH can be written in block-diagonal form:

\be
f =
\vec \phi^T
\renewcommand{\arraystretch}{1.5}
\left(
\begin{array}{cc|c}
\bar e_s &  \bar\chi_{sp} & \bar\chi_{sa} \\
\bar\chi^T_{sp} & \bar e_p & \bar\chi_{pa} \\
\hline
\bar\chi^T_{sa} & \bar\chi^T_{pa} & \bar e_a
\end{array}
\right)
\vec \phi
\ee
where the three blocks correspond to the solvent (s), polymer or protein scaffold (p), and additives or other solutes (a). The matrices $\bar e_s$, $\bar e_p$ and $\bar e_a$ are the entropic contribution associated with each class of components, while the off-diagonal matrices $\bar \chi_{ij}$ describe their mutual interactions. In particular, the lower-right block describes interactions between multicomponent biological solutes that may not themselves form the scaffold but can nevertheless mediate interactions between scaffold components. The corresponding off-diagonal blocks therefore represent phenomenological interactions between protein monomers and this diverse set of additives or solvation agents.

In this representation, the entropic contributions can be written as

\be
(\bar e_\mu)_{ij} = \delta_{ij}\frac{\ln \phi_i}{N_i} \phi^{-1}_j
\ee
with $\delta$ the Kronecker delta and $\vec \phi = (\phi_s, \vec \phi_p,\vec \phi_a)$ . 

The Schur complement \cite{zhang06}, a matrix quantity intimately related to Gaussian integration, provides a convenient mathematical representation for tracing out the fast fluctuations of the mixture. Integration over the additive degrees of freedom therefore generates an effective interaction matrix for the remaining solvent and protein variables:

\bqa
\bar f_\text{eff} =&&
\renewcommand{\arraystretch}{1.5}
\left(
\begin{array}{cc}
\bar e_s &  \bar\chi_{sp} \\
\bar\chi^T_{sp} & \bar e_p
\end{array}
\right) \nonumber\\
&-& 
\renewcommand{\arraystretch}{1.5}
\left(
\begin{array}{c}
 \bar\chi_{sa} \\
\bar\chi_{pa}
\end{array}
\right)
\times
\bar e_a^{-1}
\times
\renewcommand{\arraystretch}{1.5}
\left(
\begin{array}{cc}
\bar\chi^T_{sa} & \bar\chi^T_{pa}
\end{array}
\right)
\eqa

From this expression it is straightforward to see that the effective free energy is the Schur complement of the total free energy with respect to the fast-moving degrees of freedom.

The second term is key: although the additives have been eliminated as explicit degrees of freedom, their fluctuations leave behind an effective interaction between the remaining slow variables. In this sense, integrating out the fast degrees of freedom renormalises the protein–protein and protein–solvent interactions rather than simply removing the influence of the additives. We can assume without loss of generality that $\bar e_a$ is diagonal (additives do not interact among themselves), and so is $\bar e_a^{-1}$. Perturbations of the sort $\bar\chi_{pa}\bar e_a^{-1}\bar \chi^T_{sa}$ have four components: 

\bqa
\bar\chi_{pa}\bar e_a^{-1}\bar \chi^T_{sa} = A_{pa} \bar e_a^{-1} A_{as} &+& \\
  \frac{B_{pa} \bar e_a^{-1} A_{as}+ A_{pa} \bar e_a^{-1} B_{as} }{T} &+&  \frac{B_{pa} \bar e_a^{-1} B_{as}}{T^2} \nonumber
\eqa

Crucially, this form encodes two contributions to the free energy landscape, which encode the interplay between entropic and enthalpic effects. The temperature independent term asymptotically dominates at high temperature (cf. Eq.~\ref{eqchi}). The $1/T$ dependent term describes exothermic or endothermic mixing, depending on the involved parameters and dominates at intermediate temperatures. It is reminiscent of the empirical approach described earlier and it will either increase or decrease but eventually converge to the asymptotic value. The term that adds interesting new dynamics is the purely enthalpic one. It is worth noting that (i) it is proportional to $-T^{-2}$ and dominates at low temperature (at the core of the LCST), and that (ii) $\bar e_a^{-1}$ involves taking logarithms of normalised concentrations, which yields negative values. This means that if $B_{pa}$ and $B_{as}$ depend on temperature in opposite ways, i.e. $B_{pa}= \pm|B_{pa}|$ and $B_{as} =\mp|B_{as}|$, then the associated mixing barrier increases with temperature:

\be
\frac{d}{dT} \left( - \frac{B_{pa} \bar e_a^{-1} B_{as} }{T^2} \right) \geq 0
\ee

This behaviour (giving rise to a an increasing $\chi$ with temperature) can be interpreted as an entropic force arising from underlying electrostatic mechanisms (see Appendix). The interplay between the $1/T$ and $1/T^2$ terms will determine the nature of the reentrant phase (see Fig.~\ref{mixingbarrier}).

This temperature-dependent effective interaction can be interpreted as an emergent entropic force arising from the reorganization of the fast solvent and solute degrees of freedom in response to the slow protein variables (see Appendix). It provides a microscopic rationale, within the mean-field framework, for an LCST-like phase behavior, i.e. increasing temperature progressively strengthens the effective repulsion between the dilute and condensed states, eventually driving demixing. At sufficiently high temperatures, the competition between this temperature-dependent contribution and the other enthalpic and entropic terms can also produce non-monotonic effective interactions, providing a natural route to reentrant phase behavior

\subsubsection{Projection onto a Quasi-Binary Mixture}

The Schur-complement reduction considered above applies to the quadratic (Gaussian) approximation of the free energy, obtained by expanding the full nonlinear free energy about a chosen equilibrium state, which is well-described by the Hessian of the free energy at that point (see Appendix). This local expansion provides a natural framework for assessing the stability of the homogeneous phase and, in particular, for identifying the spinodal, where the curvature of the free energy vanishes along the unstable mode. It is important, however, to distinguish this local stability analysis from the determination of the binodal. The binodal is a global property of the nonlinear free-energy landscape, defined by coexistence between two phases of equal chemical potential and osmotic pressure, and therefore cannot in general be obtained from a purely quadratic approximation, which contains only local information around the reference state. Consequently, the Schur complement of the free energy should be interpreted here as yielding an effective interaction matrix that captures how the eliminated salt degrees of freedom modify the local protein free-energy curvature and spinodal stability. A quantitative prediction of the full binodal would instead require integrating out the salt degrees of freedom from the full nonlinear free energy before determining the resulting coexistence conditions (under the form of tangent hyperplanes).

This approach is particularly relevant, for example, to the description and programming of biomolecular condensates \cite{chen26}, which plays a critical role in polypeptide packaging by enabling dynamic and reversible condensation of protein and RNA chains into dense assemblies. In many applications, however, the detailed structure of the multidimensional free-energy landscape is not the primary quantity of interest. Rather, one seeks to describe the separation between a dilute phase, composed predominantly of solvent (water or cytoplasm), and a condensed phase, enriched in proteins, RNA, polymers, or other scaffold components. In this situation, the system can be viewed, to a good approximation, as a quasi-binary mixture between the solvent and the collection of non-solvent components treated collectively as an interacting polymeric phase. Defining the collective vector of non-solvent volume fractions as $\vec{\phi^{C}} = (\phi^{C}_{1},\phi^{C}_{2}, \dots, \phi^{C}_{N})$ such that the solvent concentration $\phi_s = 1 -\sum^N_i \phi^{C}_i$.

The relevant collective mode therefore corresponds to a change in the total non-solvent fraction, accompanied by the compensating change in solvent fraction. Projecting the Hessian onto this collective direction yields an effective curvature for the quasi-binary mixture. If this projected curvature becomes negative,

\be
\vec{\phi^{C}}^T H \vec{\phi^{C}} \leq 0
\ee
the corresponding collective fluctuation lowers the free energy, indicating that the homogeneous mixture is locally unstable with respect to demixing along that mode. Conversely, a positive curvature indicates local stability against the corresponding collective fluctuation. In this way, the multidimensional polymer–salt–solvent system can be reduced to an effective quasi-binary description while retaining the influence of the microscopic interactions and of the eliminated degrees of freedom through the effective interaction matrix.

\section{Application: In vitro LLPS for Rabies Phosphoprotein P}

\begin{figure}
\includegraphics[width=\linewidth]{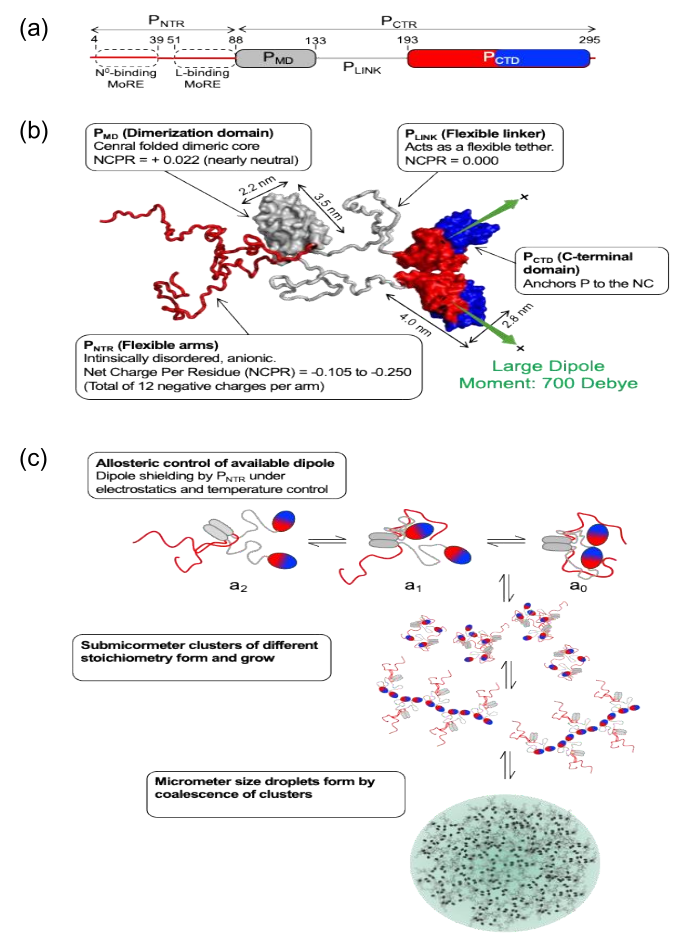}
  \caption{Architecture of RABV phosphoprotein and assembly model. {\bf (a)} Schematic
representation of the modular architecture of the rabies virus P-protein. The boxes indicate the location of folded domains, dashed boxes show the location of $N^0$- and L-binding MoREs (i.e. Molecular Recognition elements - regions that are disordered in isolation but fold or adopt a rigid conformation upon binding to a partner) and the lines show intrinsically disordered regions. {\bf (b)} Cartoon representation of one possible conformer of the RABV P dimer. The model includes the crystal structures of the dimerization domain (3 l32.pdb) and C-terminal domain (1vyi.pdb). $P_\textrm{NTR}$ are shown as red lines, dimeric $P_\textrm{MD}$ in gray surface, $P_\textrm{LINK}$ in gray lines and the $P_\textrm{CTD}$ is shown in surface colored according to its electric polarization. The green arrows show the orientation of the dipole moments. {\bf (c)} Assembly model. The upper level of the figure shows the three limiting conformational states $(a_0, a_1, a_2)$ postulated in our model. RABV P is
represented in schematically in the same color code as panel B. The second level indicates that $a_1$ and $a_2$ through their free $P_\textrm{CDT}$ assemble in multimers (dimers, trimers, multimers,… of dimers). The third level shows the formation liquid droplets by coalescence of large multimers.} 
\label{rabv} 
\end{figure}

We next evaluate the predictive power of our model by comparing its predictions with in vitro experiments on the rabies virus (RABV) phosphoprotein (P), a system that exhibits both LCST-type phase behavior and salt-dependent reentrant phase separation \cite{bouchama25}. RABV is an enveloped, negative-sense single-stranded RNA virus belonging to the \emph{Lyssavirus} genus. It causes rabies, a highly lethal neurotropic disease that is most commonly transmitted to humans through the bite of an infected animal. Once clinical symptoms appear, the disease is almost invariably fatal, accounting for an estimated 50,000–60,000 human deaths worldwide each year \cite{ribeiro24}.

During rabies virus infection, P interacts with the nucleoprotein N to drive the formation of liquid-like condensates \cite{lahaye09,nikolic16,nevers22}. These condensates promote viral replication by concentrating viral proteins and host factors while shielding viral components from cellular stress and immune surveillance. \cite{nikolic16}. Known as Negri bodies, these membraneless cytoplasmic inclusions are a hallmark of rabies virus infection in neurons and a key diagnostic feature of rabies disease. Negri bodies function as viral replication factories and display characteristic properties of liquid condensates, including fusion, dynamic exchange of components with the surrounding cytoplasm, and sensitivity to environmental conditions  \cite{nikolic17, albertini11}.

RABV P is a modular protein that contains long intrinsically disordered regions and exists natively as a stable dimer (Fig.~\ref{rabv}(a)). Its architecture comprises an intrinsically disordered N-terminal region ($P_{\text{NTR}}$), a central dimerization domain ($P_{\text{MD}}$), an intrinsically disordered linker region ($P_{\text{LINK}}$), and a folded C-terminal domain ($P_{\text{CTD}}$) characterized by a large electric dipole moment (Fig.~\ref{rabv}(b)). The long disordered $P_{\text{NTR}}$ and $P_{\text{LINK}}$ regions, connected to the central $P_{\text{MD}}$, give rise to star-like dimers with exceptional conformational flexibility, in which the two $P_{\text{CTD}}$s behave as independently mobile interaction modules (Fig.~\ref{rabv}(b)). This dynamic architecture enables P to function as a multivalent molecular scaffold, transiently bridging P molecules, recruiting unassembled monomeric N ($N^0$), which is required for the assembly of new nucleocapsids, and the viral RNA-dependent RNA polymerase L through $P_{\text{NTR}}$, while binding the N–RNA nucleocapsid template through $P_{\text{CTD}}$ and recruiting diverse cellular partners through distinct regions. Through these multiple interactions, P coordinates the assembly and organization of viral replication factories within biomolecular condensates (see Fig.~\ref{rabv}(c)).

This compartmentalization is thought to limit the accessibility of viral RNAs to cytosolic innate immune sensors such as RIG-I and MDA5 and to modulate the antiviral response by concentrating the phosphoprotein P, a multifunctional antagonist of type I interferon production and signaling \cite{scrima23,minami26,vidy05}.

\begin{figure}[b!]
\includegraphics[width=\linewidth]{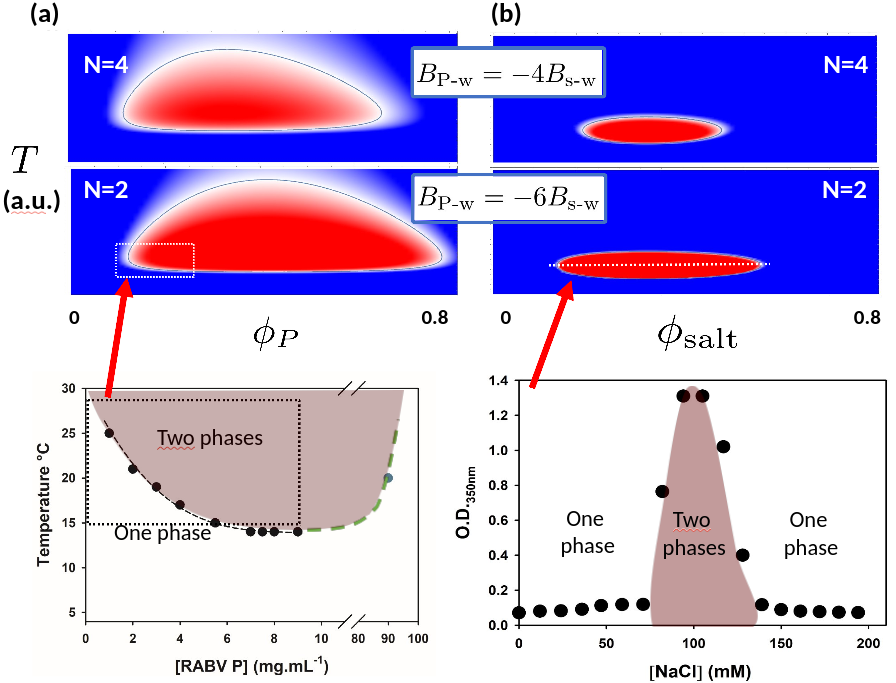}
  \caption{Analytical calculations of spinodal locii for a ternary mixture of P-protein, salt and water (temperature is in absolute units). Simulation parameters are $A_\textrm{salt-water}=A_\textrm{P-water}= 10^{-1} A_\textrm{P-salt}=10^{-2}B_\textrm{P-salt}$ and $B_\textrm{salt-water} = -10B_\textrm{P-salt} = -1$. Colour code: positive hessian $\rightarrow$ blue, negative hessian $\rightarrow$ red. The spinodal is a solid superimposed line. This exemplifies the predictive power of our theory, as we can establish different threshold ratios of $|B_\textrm{P-water}| / |B_\textrm{salt-water}|$, depending on whether the P scaffold is a dimer ($N=2$ stickers) or a tetramer ($N=4$ stickers), for the system to feature both LCST and salt-driven reentrants. \textbf{(a)} Textbook calculations of the FH barrier for this ternary mixture typically predict that the entropic components $A_{\mu\nu}$ of $\chi_{\mu\nu}$ are much smaller than $B_{\mu\nu}$. While they also predict that $B_\textrm{salt-water} \approx - B_\textrm{P-water}$. Our model indicates that no LCTS happens unless $|B_\textrm{P-water}|  \gtrapprox  5 |B_\textrm{salt-water}|$ for tetramers and slightly higher for dimers, at $\phi_\textrm{salt} = 0.01$. \textbf{(b)} Reentrants for different salt concentrations $\phi_s$ as a function of temperature for $\phi_P = 0.01$. Depending on whether the P scaffold , the reentrants exist only for above a certain value for $|B_\textrm{P-water}| / |B_\textrm{salt-water}|$. \textbf{(c)} Experimental results from Ref.~\cite{bouchama25} that show the qualitative agreement between this theory and experiment. Two leftmost correspond to an experimentally characterised LCST, whereas the two rightmost confirm a RPT with varying salt concentrations.} 
\label{simulations} 
\end{figure}

In vitro experiments \cite{bouchama25} have shown that P-proteins self-associate into coacervates, even in the absence of other agents, and feature monovalent-salt modulated reentrants. We now use the methodology outlined above to provide an interpretation of the experiment presented in Ref.~\cite{bouchama25}. The simplified model for this ternary mixture is:

\be
\bar f =
\renewcommand{\arraystretch}{1.5}
\left(
\begin{array}{cc|c}
e_\textrm{water} & \chi_{wp} & \chi_{ws} \\
\chi_{wp} & \bar e_\textrm{protein} & \chi_{ps} \\
\hline
\chi_{ws} & \chi_{ps} & e_\textrm{salt}
\end{array}
\right)
\ee
where interactions are of the form $\chi_k = A_k+B_k/T$, and should be interpreted as phenomenological free parameters (for instance encoding for charge segregation in sequences \cite{das13}). We carried out analytical calculations for this mixture, and the obtained results are presented in Fig.~\ref{simulations}. We will see below how this needs to be modified to yield the entropic LCST and reentrants. After integrating out the salt, the model is:

\bqa
f_\textrm{eff} &=& (e_w - \chi_{ws} e_s^{-1} \chi_{sw}) \phi_w \phi_w \nonumber\\
&&+ (e_p - \chi_{ps} e_s^{-1} \chi_{sp})\phi_p \phi_p \nonumber\\
&&+ (\chi_{pw} -  \chi_{ws} e_s^{-1} \chi_{ps})  \phi_p \phi_w
\eqa
where the salt variables ($\phi_s$) have been eliminated analytically, yielding an effective free-energy model that depends only on the protein and solvent variables ($\phi_p$,$\phi_w$).

In the LCST induced by RABV P, we have hypothesized that condensate assembly is driven (at least partially) by macrodipole-macrodipole interactions \cite{bouchama25}. In this framework, three potentially competing entropic contributions can be identified to explain the LCST-type behavior (see Fig.~\ref{entropicforces} in the appendix):
 
{\bf (i)} The first one is the counterion release entropy, arising from the released of counterions when the C-terminal dipolar domains come into contacts (Fig.~\ref{entropicforces}). Small monovalent ions such as Na+ or Cl- can be displaced from the vicinity of the protein upon association of the macromolecules, thereby increasing their translational degrees of freedom and contributing favourably to association.  As a result, the ions preferentially remain in the bulk solvent rather than approaching the protein surface. This produces an effective protein–salt repulsion ($B_{sp} > 0$), characteristic of a weak kosmotropic effect. The macrodipoles are then free to link up with other P dimers in solution via electrostatic interactions and a LCST emerges (see Fig.~\ref{lcst}(a)). This effect is expected to be relatively modest compared with the release of multivalent or strongly confined species, because the gain in translational entropy per released monovalent ion is limited. Nevertheless, counterion release can provide a favourable entropic contribution to the association of the PCTD domains. 

{\bf (ii)} The second one is a hydration entropy, associated with the preferential hydration of $P_{\text{NTR}}$, $P_{\text{LINK}}$ and macrodipolar $P_{\text{CTD}}$. These charged regions are surrounded by dynamic hydration shells whose organization depends on their local environment. Association of water molecules with the highly polar $P_{\text{NTR}}$, $P_{\text{LINK}}$ and $P_{\text{CTD}}$ may constrain their conformational dynamics relative to their freely solvated states in bulk water. From an entropic perspective, it is therefore unfavourable to maintain hydration shells in constrained positions near charged regions and macrodipoles.  This gives rise to a weakly attractive interaction between the $P_{\text{CTD}}$ domains, corresponding to: $B_{pa} > 0$ (see Fig.~\ref{entropicforces}). In this line, the competition between the favourable protein–protein interactions and the unfavourable organization of ions and water provides a possible microscopic origin for salt-dependent reentrant phase behaviour. 

{\bf (iii)} The third is a conformational entropy of the protein, arising from salt-dependent changes in the conformational ensemble of P. We hypothesize that the negatively charged, intrinsically disordered N-terminal regions may interact electrostatically with the positively charged pole of the C-terminal macrodipole, thereby partially shielding the macrodipole and restricting the conformational freedom of the N-terminal and $P_{\text{LINK}}$ regions \cite{bouchama25}. Addition of salt could weaken these intramolecular electrostatic interactions, releasing these conformational restraints. Because these regions are intrinsically disordered, their release would increase the number of accessible conformational microstates and thus the conformational entropy of the protein. At the same time, exposure of the C-terminal macrodipoles would allow stronger intermolecular dipole–dipole interactions and thereby favor protein assembly. This mechanism therefore couples an entropically favorable increase in protein conformational freedom to an enthalpically favorable intermolecular interaction. Together, these competing enthalpic and entropic contributions could provide a molecular basis for the observed temperature dependence of phase separation

Finally, the model was also able to capture the reentrant phase behaviour as a function of salt concentration (see Fig.~\ref{simulations}(b)). We calculated analytically the spinodal locus for a range of parameters and found that it is qualitatively consistent with the observations in Ref.~\cite{bouchama25}. Crucially, the threshold enthalpic barrier for the RPT to exist depends on the number of sticky patches $N$. Therefore, $N$ can be understood as a free parameter (depending on the oligomerasation of the proteins) which needs to be fitted experimentally. Indeed, one would expect that the possible combinations of $\chi$ and $N$ can be experimentally restricted by combining data from independent experiments on LCST and salt-based RPT. 

\section{Discussion and Outlook}

We have presented a mean-field model for biological condensates, inspired by matrix formulation of FH theory, that captures both LCST behaviour arising from entropic effects and reentrant phase behaviour driven enthalpic interactions.  

The main insight of this work is that integrating out fast degrees of freedom, such as small ligands, ions or other co-solutes, has a natural interpretation within the matrix formalism through the Schur complement of a quadratic free energy, a mathematical operation closely related to Gaussian integration. This framework reveals how effective entropic forces can emerge from underlying enthalpic interactions by modifying the number of available microstates. In this respect, the mechanisms bears similarities to the classical ``salting-out" effect induced by kosmotropic ions. More importantly, however, the approach is sufficiently general to describe a broad range of biological systems undergoing LLPS, including multicomponent condensates, and that provides a unified framework for incorporating more general mechanisms leading to emergent entropic interactions. 

We tested the applicability of the model by reproducing the qualitative features of recent in vitro experiments on phase separation of RABV P protein over a limited range of salt concentrations. More generally, the formalism can be extended to describe more complex multicomponent models, including those in which transitions between homotypic to heterotypic LLPS play a dominant role \cite{rana24,welles24,onuchi19,patel17}.

The approach proposed here is primarily phenomenological, and its predictive power remains to be rigorously tested.  A quantitative validation would require fitting several experimentally determined parameters, including the effective size of the proteins(and the corresponding number of ``sticky patches"), as well as the entropic and enthalpic contributions associated with salt and dipolar interactions. The latter may provide a way toward recovering Pappu's $\kappa$ parameter, which quantifies charge segregation in polyampholyte \cite{das13}). Despite these limitations, it is remarkable that the model can qualitatively reproduce both reentrant phase separation as a function of salt concentrations and the emergence of an LCST under the simple assumption that protein-water entropic interactions are regulated by the presence of salt. We emphasize that the agreement with the experimental data is currently qualitative rather than quantitative. As additional experimental measurements become available, we intend to further refine the model and increase its predictive power. Furthermore, the formalism is readily suited to automated parameter fitting, a research  avenue that we intend to pursue in future work. 

Another particularly intriguing aspect of the present framework is the connection between Onsager reciprocal relations \cite{onsager31} and enthalpic interactions between different species, as explained in the appendix. We speculate that this might find interesting applications in coacervate programming, in which one tries elicit useful responses in certain variables by controlling conjugate parameters \cite{chen26}.

\clearpage
\onecolumngrid

\section*{A1: Schur Complement and Entropic Forces}

Our approach distinguishes between two classes of dynamics: {\bf slow degrees of freedom}, which describe the protein concentration and the collective organization of the protein network, including condensate structure, and {\bf fast degrees of freedom}, which describe salt ions, water molecules, and potentially other small co-solutes that equilibrate on time scales much shorter than those of the protein degrees of freedom. We further assume that, close to equilibrium, fluctuations of these variables can be described within the Gaussian approximation, i.e. quadratic around a minimum. This separation of time scales provides the basis for integrating out the fast degrees of freedom and deriving an effective free-energy description of the slow protein variables.

The Schur complement provides an exact algebraic formulation of bath-mediated mean-field theory. We saw how this construction can provide a way of obtaining a LCST via effective attractive interactions mediated by small molecules interspersed amongst the multimers. We will now briefly relate this approach to the internal logic of effective theories and in particular its relation to the Gaussian Integration. Consider the joint Gaussian distribution:

\bqa
p(x_1, x_2) &\propto& \exp \left( -\frac{1}{2} \begin{bmatrix} x_1^T & x_2^T \end{bmatrix} \begin{bmatrix} \Lambda_{11} & \Lambda_{12} \\ \Lambda_{21} & \Lambda_{22} \end{bmatrix} \begin{bmatrix} x_1 \\ x_2 \end{bmatrix} \right)\nonumber \\
 &\propto& \exp \left( -\frac{1}{2} \left( x_1^T \Lambda_{11} x_1 + x_1^T \Lambda_{12} x_2 + x_2^T \Lambda_{21} x_1 + x_2^T \Lambda_{22} x_2 \right) \right)
\eqa
with $\Lambda_{21}^T = \Lambda_{12}$. The block $\Lambda_{11}$ describes the direct protein-protein interactions. In the FH framework, it accounts, in a coarse-grained sense, for self-interactions, i.e. protein compressibility and the direct contribution of protein mixing. The block  $\Lambda_{22}$ describes fluctuations of the fast degrees of freedom (salts and/or other co-solutes); in the absence of coupling to the protein, it characterizes the intrinsic response of the salt bath. Finally, the off-diagonal $\Lambda_{12}$  describes the coupling between protein and salt fluctuations. In the FH formulation, this coupling is embodied by the protein–salt interaction parameter $B_{ps}$, or, more generally, by the corresponding protein–salt interaction matrix.

The marginal distribution of $x_1$:

\be
p(x_1) = \int_{-\infty}^{\infty} p(x_1, x_2) \, dx_2  \propto \exp \left( -\frac{1}{2} x_1^T \Lambda_{11} x_1 \right) \int_{-\infty}^{\infty} \exp \left( -\frac{1}{2} \left( x_2^T \Lambda_{22} x_2 + 2 x_2^T \Lambda_{21} x_1 \right) \right) \, dx_2
\ee

Gaussian integrals are easier to solve by centering the integrand. Note that minimising the exponent gives the equilibrium condition $\nabla_{x_2} \left( x_2^T \Lambda_{22} x_2 + 2 x_2^T \Lambda_{21} x_1 \right) = 2 \Lambda_{22} x_2 + 2 \Lambda_{21} x_1 = 0 \rightarrow x_2 = -\Lambda_{22}^{-1} \Lambda_{21} x_1$, which tells us that we can interpret quadratic fluctuations of $x_2 = x^* + \delta x_2$ as happening around the minimum $x^*_2 = -\Lambda_{22}^{-1} \Lambda_{21} x_1$. This yields:

\bqa
p(x_1) \propto &&\exp \left( -\frac{1}{2} x_1^T \Lambda_{11} x_1 \right) \int_{-\infty}^{\infty} \exp \left( -\frac{1}{2} \left( (x^* + \delta x_2)^T \Lambda_{22} (x^* + \delta x_2) + 2 (x^* + \delta x_2)^T \Lambda_{21} x_1 \right) \right) \, dx_2 \nonumber\\
= && \exp \left( -\frac{1}{2} x_1^T \Lambda_{11} x_1 \right) \cdot \exp \left( \frac{1}{2} x_1^T \Lambda_{12} \Lambda_{22}^{-1} \Lambda_{21} x_1 \right) \times \int_{-\infty}^{\infty} \exp \left( -\frac{1}{2}  \delta x_2^T \Lambda_{22} \delta x_2 \right) \, d(\delta x_2)\nonumber \\
\propto && \exp \left( -\frac{1}{2} x_1^T \left( \Lambda_{11} - \Lambda_{12} \Lambda_{22}^{-1} \Lambda_{21} \right)x_1 \right)
\eqa
as the Gaussian integral reduces to a normalisation factor. In the end we get that the resulting distribution, up to this normalisation factor, for the $x_1$ variable comes given by Schur's complement:

\be
\boxed{
\Lambda_{\text{eff}} = \Lambda_{11} - \Lambda_{12} \Lambda_{22}^{-1} \Lambda_{21}
}
\ee

Originally, proteins interacted directly through $\Lambda_{11}$. After eliminating the salt component, proteins appear to interact through $\Lambda_{11} - \Lambda_{12} \Lambda_{22}^{-1} \Lambda_{21}$. Thus, salt creates an effective interaction between proteins independently of their direct interaction (even if 11=0). 

The correction term $\Lambda_{12} \Lambda_{22}^{-1} \Lambda_{21}$ is the effective interaction induced by the fast fluctuations which were integrated out. In the context of biological solutions, this identifies mean-field renormalisation with elimination of Gaussian degrees of freedom whose dynamics happens on a much faster scale than that of the scaffolds.

\subsection{Interpretation as an Entropic Force}

We now explain why the effective interaction mediated by the Gaussian fluctuations integrated out has an entropic interpretation. Enthalpic processes typically change the number of available microstates (see for instance \cite{yu23} in the case of tethered IDPs). If the solution is perturbed in such a way that the number of microstates is increased, an effective entropic force arises to restore equilibrium. The probability of a configuration $x$ is related to the number of microstates $\Omega(x)$:

\be
p(x) \propto \Omega(x) = \exp\left(\frac{S(x)}{k_B}\right)
\ee
where $S(x)$ is the entropy associated to the macrostate to which the microstate belongs. Assuming that the system is in near-equilibrium, i.e. it maximises entropy, small perturbations mediated by the fast degrees of freedom will result in a deviation from the maximum possible entropy:

\be
S_{\text{marginal}}(x_1) = S_\textrm{MAX} - \frac{1}{2} k_B x_1^T \left( \Lambda_{11} - \Lambda_{12} \Lambda_{22}^{-1} \Lambda_{21} \right) x_1
\ee

Depending on the sign of the mediated interaction term $- \Lambda_{12} \Lambda_{22}^{-1} \Lambda_{21} $, the equilibrium of the renormalised system may differ from the original one. An entropic force due to the variation of parameter $X$ is of the form (\cite{onsager31}):

\be
F_{\text{entropic}} = \frac{d}{dX} S
\ee
which aims at restoring the maximum entropy of the system (i.e. maximising the number of available microstates).  In other words, if an enthalpic interaction increases the number of available microstates, it will give rise to an entropic force that until the entropy is maximised.

We have seen in the case of Rabies P-protein that the effective force is dominated by the term $-1/T^2$ at low temperatures, which means that increasing temperature will result in an entropic increase of the mixing barrier. 

Indeed in our model,

\be
\chi_{eff}(T)= \chi_0 + \Delta\chi(T)
\ee

\be
\Delta\chi(T) \propto-\Lambda_{12} \Lambda_{22}^{-1} \Lambda_{21} 
\ee

If this correction grows with temperature (for example through a dominant contribution arising from the temperature dependence of the coupling matrices), then increasing T makes the effective protein-protein attraction stronger. This raises the free-energy penalty for mixing and can drive an LCST transition. The ``entropic" character comes from the fact that this effective interaction originates from integrating out the fast solvent/salt degrees of freedom rather than from a direct protein-protein energetic interaction.

\begin{figure}[h!]
\includegraphics[scale=0.25]{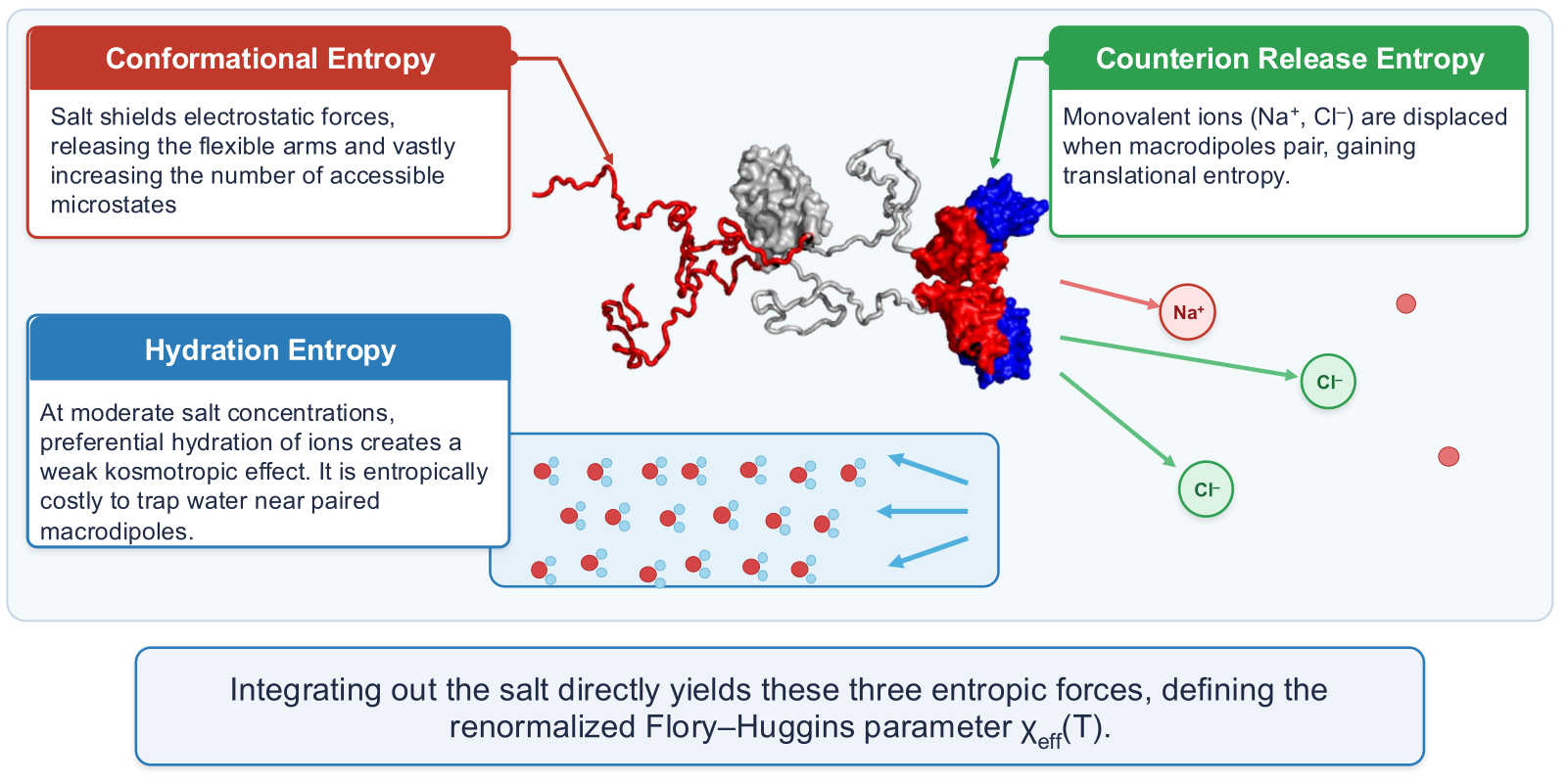}
  \caption{Anatomy of an emergent entropic force. This shows competing entropic contributions that explain the emergence of a LCST-type behavior in the RABV P system. For a given salt concentration, the protein conformational energy is maximised if the intrinsically disordered region is energetically allowed to unfold. This can happen, for instance, because at moderate concentrations the macrodipoles are starved of water molecules by the salt ions. This in turn, increases the traslational energy of water molecules, increasing the number of microstates.} 
\label{entropicforces} 
\end{figure}

\section*{A2: Kramer's Escape Rate and Matrix Determinants}

\begin{figure}[b!]
\includegraphics[scale=0.3]{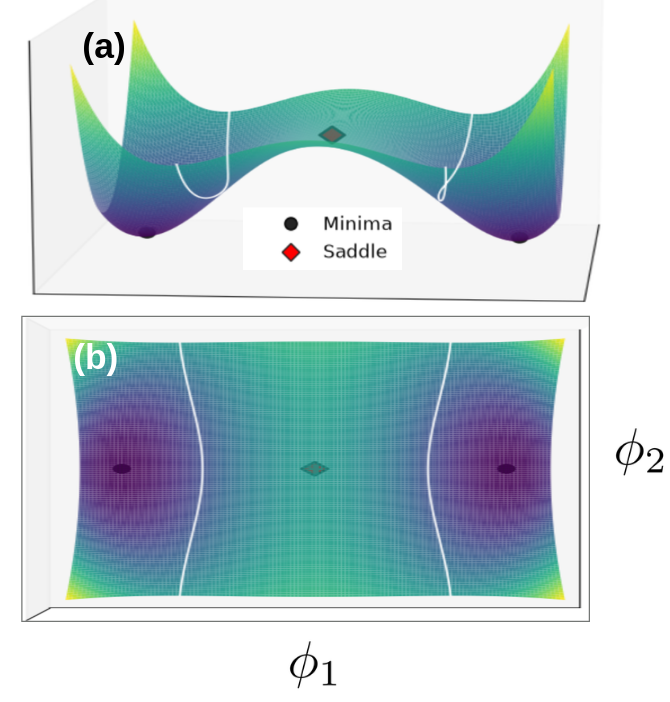}
  \caption{Generalisation of the free energy landscape to two dimensions.{\bf (a)}A two-dimensional entropic parabola, for two polymer species $\phi_1$ and $\phi_2$ is separated into wells by an enthalpic barrier ($\sum^{1,2,\textrm{solvent}}_i\phi_i = 1$). In this case, the two minima correspond to a positive-definite Hessian while the saddle corresponds to a point with at least one negative eigenvalue (see appendix).{\bf (b)} High-angle view of the same landscape. White curves denote the spinodal condition, i.e. $det H = 0$, which is the hallmark of a transition from convex, stable minima to an unstable region.} 
\label{2dlandscape} 
\end{figure}

\begin{figure}[h!]
\includegraphics[scale=0.25]{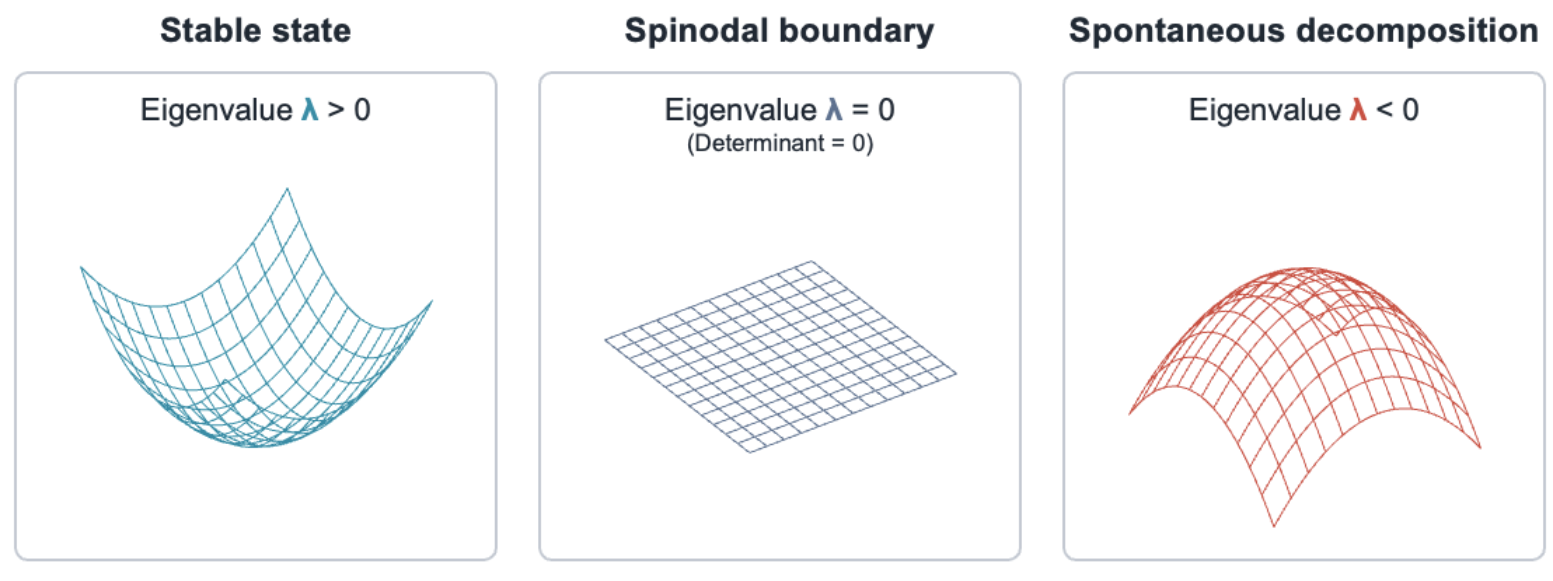}
  \caption{Stability conditions for local spinodal condition, in terms of the curvature of the free energy. If all eigenvalues are positive, the free energy landscape curves upward, and the concentration is stable. If eigenvalues are zero, it takes no energy to initiate a transition along the flat (constant energy) directions. It also marks a transition from a stable regime (metastable binodal) from an unstable regime (spinodal). If some eigenvalues are negative, the free energy landscape curves downward and any perturbation will move the concentration away from the stable point (maximum). } 
\label{eigenvalues} 
\end{figure}

As is frequently done in statistical physics, free energy minimisation can be understood as the (overdamped) motion of a dissipative point-like particle evolving on a free-energy landscape. In a one dimensional system, such as the canonical FH model, stable and metastable states correspond to local minima of the free energy, while the transition between them involves crossing a free-energy barrier. In this picture, phase separation from a metastable state can be described as an activated escape process, for which Kramers' theory provides the corresponding transition rate \cite{altland10} (see Figs. \ref{2dlandscape} and \ref{eigenvalues}).

To compute the tunnel rate of a multidimensional potential $U(\mathbf{x})$, one can approximate it  up to second order using a Taylor expansion (matricial expansion) around the critical points \cite{hanggi90}. The local geometry is dictated by the Hessian matrix $H$, defined element-wise as:

\be
H_{ij} = \frac{\partial^2 U}{\partial x_i \partial x_j}
\ee

At a local minimum ($\mathbf{x}_\textrm{min}$), the Hessian matrix $H_\textrm{min}$ is positive-definite. All its eigenvalues $\lambda_i^{(\textrm{min})} > 0$ for $i = 1, \dots, d$. The local potential is approximated as:
\be
U(\mathbf{x}) \approx U(\mathbf{x}_\textrm{min}) + \frac{1}{2} (\mathbf{x} - \mathbf{x}_\textrm{min})^T H_a (\mathbf{x} - \mathbf{x}_\textrm{min})
\ee

At a saddle point ($\mathbf{x}_\textrm{saddle}$), the transition state is characterized by an indefinite Hessian matrix $H_\textrm{saddle}$. Specifically, it possesses at least one negative eigenvalue ($-\lambda_1^{(\textrm{saddle})}$, where $\lambda_1^{(\textrm{saddle})} > 0$), corresponding to the unstable direction (the reaction coordinate, as it were). The remaining eigenvalues $\lambda_i^{(\textrm{saddle})}$ are positive, corresponding to stable orthogonal modes that form the activation bottleneck.

The spinodal condition amounts to the transition between positive values for some of the Hessian eigenvalues, i.e. $\lambda_k=0$, and reduces trivially to:

\be
\det  H = 0
\ee

Evaluating the partition functions via multivariate Gaussian integration yields the generalized multidimensional Kramer's tunneling rate \cite{hanggi90}:

\be
\Gamma \propto \sqrt{\frac{\det H_\textrm{min}}{|\det H_\textrm{saddle}|}} \exp\left( -\frac{\Delta U}{k_B T} \right)
\ee

where $\det H_\textrm{min} = \prod_{i=1}^d \lambda_i^{(\textrm{min})}$ measure the number of microstates available at the bottom of the minimum. Similarly, $|\det H_\textrm{saddle}| = \lambda_1^{(\textrm{saddle})} \prod_{i=2}^d \lambda_i^{(\textrm{saddle})}$ captures the amount of the stable modes that can pass through the saddle point. If one allows the concentration to be a dynamical variable in space, this method can be used to analise spinodal behaviour.

\end{document}